\documentclass{article}

\PassOptionsToPackage{numbers}{natbib}

 \usepackage[preprint]{neurips_2025}

\usepackage[utf8]{inputenc} 
\usepackage[T1]{fontenc}    
\usepackage{hyperref}       
\usepackage{url}            
\usepackage{booktabs}       
\usepackage{amsfonts}       
\usepackage{nicefrac}       
\usepackage{microtype}      
\usepackage{xcolor}         

\usepackage[acronym]{glossaries}  
\usepackage{graphicx}  
\usepackage{subcaption}  
\usepackage{wrapfig}   
\usepackage{tabularx}  
\usepackage{multirow}  
\usepackage{booktabs}  
\usepackage{rotating, siunitx}  

\newacronym{AI}{AI}{Artificial Intelligence}
\newacronym{NLP}{NLP}{Natural Language Processing}
\newacronym{LLM}{LLM}{Large Language Model}
\newacronym{ICL}{ICL}{in-context learning}
\newacronym{SLO}{SLO}{Service Level Objective}
\newacronym{LoRA}{LoRA}{Low-Rank Adaptation}
\newacronym{QoS}{QoS}{Quality of Service}
\newacronym{TPOT}{TPOT}{Time Per Output Token}
\newacronym{ITL}{ITL}{Inter-Token Latency}
\newacronym{TTFT}{TTFT}{Time To First Token}
\newacronym{FCFS}{FCFS}{First Come First Served}
\newacronym{LRU}{LRU}{Least Recently Used}
\newacronym{SMAPE}{SMAPE}{Symmetric Mean Absolute Percentage Error}

\title{A Data-driven ML Approach for Maximizing Performance in LLM-Adapter Serving}

\author{
\textbf{Ferran Agulló}\textsuperscript{1,2} \quad
\textbf{Joan Oliveras}\textsuperscript{1,2} \quad
\textbf{Chen Wang}\textsuperscript{3} \quad
\textbf{Alberto Gutierrez-Torre}\textsuperscript{1} \\  
\textbf{Olivier Tardieu}\textsuperscript{3} \quad
\textbf{Alaa Youssef}\textsuperscript{3} \quad
\textbf{Jordi Torres}\textsuperscript{1,2} \quad
\textbf{Josep Ll. Berral}\textsuperscript{1,2} \\
\textsuperscript{1}Barcelona Supercomputing Center (BSC), Spain \\
\textsuperscript{2}Universitat Politècnica de Catalunya - BarcelonaTech (UPC), Spain \\
\textsuperscript{3}IBM Research, USA \\
\texttt{\{ferran.agullo, joan.oliveras, alberto.gutierrez, jordi.torres\}@bsc.es} \\
\texttt{josep.ll.berral@upc.edu} \quad
\texttt{\{chen.wang1, tardieu, asyousse\}@ibm.com}
}

\begin{document}

\maketitle

\begin{abstract}
  With the rapid adoption of Large Language Models (LLMs), LLM-adapters have become increasingly common, providing lightweight specialization of large-scale models. Serving hundreds or thousands of these adapters on a single GPU allows request aggregation, increasing throughput, but may also cause request starvation if GPU memory limits are exceeded. To address this issue, this study focuses on determining the joint configuration of concurrent and parallel adapters that maximizes GPU throughput without inducing starvation, given heterogeneous adapter and traffic properties. We propose a data-driven ML approach leveraging interpretable models to tackle this caching problem and introduce the first Digital Twin capable of reproducing an LLM-adapter serving system, enabling efficient training data generation. Experiments with the vLLM framework and LoRA adapters show that the Digital Twin reproduces throughput within 5.1\% of real results, while the ML approach predicts optimal numbers of concurrent and parallel adapters with an error of at most 7.2\% under heterogeneous, real-world workloads. The code is publicly available at \url{https://github.com/FerranAgulloLopez/GPULLMAdapterOptimization}.
\end{abstract}

\section{Introduction}
With the rapid advancement and widespread adoption of \Glspl{LLM}, the demand for \Gls{LLM}-adapters has grown significantly. While \Glspl{LLM} are large-scale models trained to achieve strong performance across diverse language tasks, adapters specialize this general knowledge to concrete applications through lightweight parameter additions~\cite{hu2022lora, houlsby2019parameter, liu2022few, guo2020parameter}. Their compact size enables serving systems to host hundreds or even thousands of adapters on a single GPU~\cite{sheng2024slora, bruel2024compress}, thereby increasing throughput by aggregating requests from many adapters. However, excessive concurrency can reach a critical threshold at which request starvation arises, as the system becomes unable to process incoming requests within available GPU memory limits. This saturation threshold is governed by the interaction of adapter size, request length, and adapter arrival rate, which together determine the degree of concurrency needed to reach the device limits. Request length and adapter size affect the per-request memory usage, while adapter size and arrival rate jointly determine the memory demand per adapter.

Moreover, GPU memory capacity is often insufficient to hold all adapter weights, and only those loaded into device memory can actively process requests. To address this, systems dynamically swap adapters between storage and GPU memory. Adapters resident on the GPU execute requests in parallel with other loaded adapters~\cite{chen2024punica}, while swapping enables concurrent execution across those that cannot fit. In some cases, limiting the number of simultaneously loaded adapters—even when memory is sufficient\textemdash can improve throughput by leaving more GPU memory available for request processing. Conversely, loading too few may prevent loaded adapters from fully utilizing the available device memory. Thus, selecting the optimal number of adapters to process in parallel is a key server configuration parameter, that also impacts throughput and the saturation threshold.

Building on the above, this study addresses the following problem: \textbf{Given a set of adapters to serve with specified heterogeneous adapter sizes, arrival rates, and request lengths, determine the joint configuration of concurrent and parallel adapters that maximizes a GPU’s achievable throughput without inducing request starvation}. We refer to this formulation as the adapter caching problem.

While prior research has investigated the optimization of \Gls{LLM}-adapter serving through kernel-level enhancements~\cite{chen2024punica}, memory management techniques~\cite{sheng2024slora}, and scheduling strategies~\cite{iliakopoulou2024chameleon}, this specific adapter caching problem remains largely under-explored. The most closely related work, dLoRA~\cite{wu2024dlora}, employs a greedy, heuristic-driven algorithm to proactively determine how many adapters to serve per GPU in a multi-device system. Advancing on this direction, our work makes two key contributions: (1) we introduce the usage of interpretable ML models to estimate the optimal joint configuration of the adapter caching problem, maximizing single-GPU throughput while preventing request starvation; and (2) we develop the first Digital Twin capable of accurately reproducing the behavior of an \Gls{LLM}-adapter serving system, enabling the timely generation of synthetic data required to train the ML models. Together, these two contributions constitute the so-called data-driven ML approach. We work with the widely adopted vLLM framework~\cite{kwon2023efficient}, in conjunction with LoRA adapters~\cite{hu2022lora}. In vLLM, the maximum number of parallel adapters in GPU is statically defined at server startup as a fixed number of adapter slots; we adopt this terminology from this point onward.

\section{Illustrating the adapter caching problem}\label{sec - Background}
\begin{figure}
    \centering
    \includegraphics[width=\linewidth]{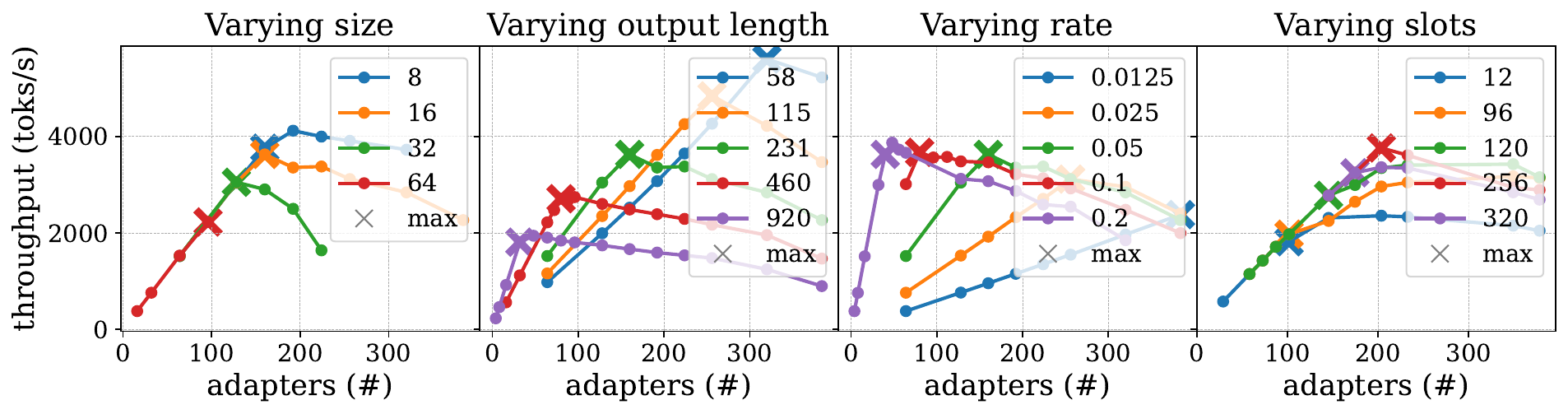}
    \caption{Throughput evolution with the number of concurrent adapters when varying adapter sizes, adapter request rates, request output lengths, and adapter slots. The results are shown for Llama-2-7B~\cite{touvron2023llama2openfoundation} and a public adapter~\cite{llama-2-7b-sql-lora-test}, using by default rate 0.05 reqs/s, adapter size 8, 250 input tokens, and 231 output tokens. The \textit{max} crosses denote the targeted optimal of the adapter caching problem.}
    \label{fig:performance_analysis-without_offloading_variation}
\end{figure}

Figure~\ref{fig:performance_analysis-without_offloading_variation} provides a visual illustration of the adapter caching problem. It depicts throughput (y-axis) as a function of the number of concurrently served adapters (x-axis) across different homogenous scenarios. Each curve exhibits two distinct regimes: initially, throughput increases proportionally with the number of adapters, as the system can accommodate the additional requests. Beyond a certain threshold, however, throughput growth slows or declines, reflecting the system’s inability to process excess requests in time, leading to request starvation. The number of concurrent adapters at the transition between these regimes, along with the adapter slots that achieve the best performance (rightmost figure), define the joint configuration targeted in the adapter caching problem for every scenario, which maximizes throughput while avoiding starvation. Practically, we identify this point as the highest measured throughput that remains above 90\% of the total incoming token rate. As noted earlier and visible in the figure, this optimal point is highly dependent on the combination of adapter size, request length, and arrival rate, with even greater variation expected in scenarios with heterogeneous characteristics, which represents the real-world behaviour tested in this study.

\section{Data-driven ML method}
Rather than relying on heuristics, we directly employ ML models to solve the adapter caching problem. We evaluate fast, lightweight, and interpretable models that enable rapid, low-resource predictions suitable for production deployment, while producing outputs that are easily interpretable and allow extraction of simple decision rules. Specifically, we employ three types of linear models\textemdash standard linear regression (LinearRegression), Bayesian Ridge regression (BayesianRidge), and Partial Least Squares Regression (PLSRegression)\textemdash, as well as, three types of decision tree-based models: the default Random Forest Regressor (RFRegressor), RuleFitRegressor~\cite{d977e822-5d0a-3b75-87ec-6aacda2e4aa4} and FIGSRegressor~\cite{tan2023fast}. All models come from the Python libraries scikit-learn~\cite{scikit-learn} and imodels~\cite{imodels2021}. The inputs for these models consist on the sizes and rates of the adapters to be served in the given scenario, encoded via the minimum, maximum, mean, and standard deviation of all values. Request lengths are fixed using a cleaned ShareGPT dataset version~\cite{ShareGPT_Vicuna_unfiltered}, retaining their heterogeneous input/output lengths. For each model type, we create two instances: one to predict the optimal number of concurrent adapters, and another to predict the optimal number of adapter slots. Ground truth for both outputs is obtained by systematically recreating the scenario defined by the inputs, in a manner analogous to Figure~\ref{fig:performance_analysis-without_offloading_variation}, evaluating a representative set of combinations of concurrent adapters and adapter slots, and selecting the one obtaining the highest throughput that does not induce request starvation. 

Due to the significant resource and time consumption of \Gls{LLM} serving benchmarking in performing these kind of searches, we introduce the first Digital Twin (DT) of an \Gls{LLM}-adapter serving system, specifically targeting the tested vLLM framework. Unlike a traditional simulator, this Digital Twin emulates the system’s state and behavior across key execution steps. Although operating offline, it reproduces scheduling, memory allocation, adapter loading, and model forward pass along a simulated timeline reflecting real-time execution. We use predictive performance models to estimate the latency of each step (expanded in Appendix~\ref{Appendix - Digital Twin extended description}). These performance models are constructed from simplifications or modifications of prior work~\cite{zhang2023shepherd, 10.1145/3341301.3359658, li2024caraserve} and are formalized in Equation~\ref{eq:digital_twin-latency} (novelty discussed in Appendix~\ref{Appendix - Novelty in Digital Twin predictive performance}). For simplicity, the predictive models are specific to each LLM–hardware combination, meaning a separate model must be trained for every combination to be evaluated.

\begin{equation}\label{eq:digital_twin-latency}
\begin{aligned}
    &\text{Latency of the backbone model forward pass} = K_4R_{running} + K_5 \\
    &\text{Latency of the adapters forward pass overhead} = K_6A_{running} + K_7 \\ 
    &\text{Latency of the scheduler} = K_1R_{running} + K_2R_{waiting} + K_3R_{waiting}(G/N)\\
    &\text{Latency for loading adapters} = \text{simple dictionary created from benchmark data} \\ 
\end{aligned}
\end{equation}
where all constants \(K_x\) are calibrated using benchmarking data with a non-linear least squares fitting (\textit{curve\_fit} method of SciPy python package~\cite{2020SciPy-NMeth}), and where \(R_{running}\), \(R_{waiting}\), \(A_{running}\), \(G\) and \(N\) denote the number of running requests, waiting requests, parallel adapters, adapter slots, and concurrent adapters, respectively.

\section{Results}
\textbf{Setup.} We use Llama-3.1-8B-Instruct~\cite{grattafiori2024llama3herdmodels} and Qwen2.5-7B-Instruct~\cite{qwen2025qwen25technicalreport} with LoRA adapters derived from two HuggingFace adapters~\cite{llama-3.1-8B-Instruct-Finance-lora-adapter, flowertune-medical-lora-qwen2.5-7b-instruct}. All experiments are conducted on a node equipped with a NVIDIA Hopper H100 (64GB HBM2), 128GB RAM memory, and 20 CPU cores.

\subsection{Modelling results}
We evaluate the Digital Twin by comparing its simulated scenarios against real-system benchmark results, as summarized in Table~\ref{table:evaluation_table}. We average similarity results under high-rank scenarios\textemdash equally considering input adapters of ranks 8, 16, and 32\textemdash and medium- to low-rank scenarios, only considering ranks of 8 and 16. For adapter rates, we evaluate high-rate scenarios (0.2, 0.1, 0.05) and low-rate scenarios (0.025, 0.0125, 0.00625). For each scenario, we simulate one hour of serving and evaluate performance metrics across a range of served adapters (8–384) and adapter slots (8–384). Overall, the Digital Twin closely reproduces real-system behavior, particularly for throughput and \Gls{ITL}, with maximum errors of 5.1\% and 9.6\%, respectively. \Gls{TTFT} exhibits a higher average error of 17.9\%. In terms of execution efficiency, the DT achieves up to a 90× speedup relative to the full one-hour being simulated, though there remains substantial room for further improvement. Resource consumption is minimal: the DT runs without a GPU, utilizes at most a single CPU core, and requires approximately 200MB of RAM (expanded results in Appendix~\ref{Appendix - Digital Twin extra results}).

\begin{table}[!htb]
    \renewcommand{\arraystretch}{1.2}
    \centering
    \sisetup{
        table-format=2.2,
        detect-weight=true,
        detect-inline-weight=math
    }
    \begin{tabularx}{\columnwidth}{>{\centering\arraybackslash}m{1.2cm} | *{6}{>{\centering\arraybackslash}X} | *{4}{>{\centering\arraybackslash}X}}
        \toprule
        & \multicolumn{6}{c}{\textbf{Digital Twin}} & \multicolumn{4}{c}{\textbf{ML model}} \\
        \toprule
        \textbf{Model} & Throu. (\%) & ITL (\%) & TTFT (\%) & Time (s) & CPU (\%) & Mem (\%) & Type & Con. Adap. (\%) & Adap. Slots (\%) & Time (ms) \\
        \midrule
        \multirow{3}{*}{\rotatebox{90}{Llama}}
            & \multirow{3}{*}{4.2} & \multirow{3}{*}{9.6} & \multirow{3}{*}{17.4} & \multirow{3}{*}{39.5} & \multirow{3}{*}{90.1} & \multirow{3}{*}{210} & Linear & 40.6 & 17.4 & 0.04 \\
            & & & & & & & Tree & 0.1 & 6.7 & 0.13 \\
            & & & & & & & \textit{Tree**} & 3.7 & 10.9 & 0.10 \\
        \midrule
        \multirow{3}{*}{\rotatebox{90}{Qwen}}
            & \multirow{3}{*}{5.1} & \multirow{3}{*}{9.1} & \multirow{3}{*}{17.9} & \multirow{3}{*}{41.4} & \multirow{3}{*}{90.7} & \multirow{3}{*}{211} & Linear & 38.6 & 23.6 & 0.03 \\
            & & & & & & & Tree & 1.0 & 7.2 & 0.15 \\
            & & & & & & & \textit{Tree**} & 1.0 & 12.4 & 0.09 \\
        \bottomrule
    \end{tabularx}
    \caption{Final results comparing Digital Twin and ML models against benchmark values. Similarity is measured using the SMAPE metric (lower values indicate closer alignment). Execution time is reported for both approaches, and resource usage is also provided for the DT results. For the ML models, we report the best results from both Linear-based and Tree-based model types (expanded in Appendix~\ref{Appendix - ML model extra results}), along with results from the interpretability solution under \textit{Tree**}.}
    \label{table:evaluation_table}
\end{table}

\subsection{Predicting optimal joint configuration}
We evaluate the ML model in identifying the joint configuration of the adapter caching problem. The evaluation focuses on scenarios with more highly heterogeneous adapter rates and sizes than the previous section, reflecting real-world conditions. Specifically, for adapter rates, we consider input adapters with rates from all combinations of three values drawn from the set [3.2, 1.6, 0.8, 0.4, 0.1, 0.05, 0.025, 0.0125, 0.00625, 0.003125]. Similarly, for adapter sizes, all possible three values are considered from the set [8, 16, 32]. For each size-rate mix, we evaluate all combinations from a range of concurrent adapter counts (8–384) and adapter slot counts (8–384) to establish the ground truth for these two ML outputs. Of these combinations, 99\% are simulated using the DT to generate the training and validation dataset (87,198 runs), while only the remaining 1\% is reserved for testing with the real system (882 runs)\textemdash given the resource- and time-intensive nature of \Gls{LLM} benchmarking. During training, we employ 5-fold cross-validation and perform hyperparameter optimization using the HalvingGridSearchCV method from scikit-learn~\cite{scikit-learn}. The right portion of Table~\ref{table:evaluation_table} summarizes the results on the testing set. Linear models fail to capture the complexity of the task, whereas tree-based models generate predictions closely aligned with actual values\textemdash achieving a maximum average error of 1.0\% for the number of concurrent adapters in both models. Prediction error for adapter slots remains higher, with a maximum of 7.2\%, suggesting room for further improvement. With an average inference time of at most 0.15 ms across all tree-based models, they are well-suited for production deployment. Additionally, for both ML outputs, we compress the best-performing tree-based model into a single, shallow tree that can be represented as fewer than 30 highly interpretable rules of the form “condition 1 AND condition 2 … → RESULT”. Despite this reduction, the tree maintains competitive performance (Table~\ref{table:evaluation_table}, row \textit{Tree**}) and provides a highly interpretable decision-making framework suitable for production deployment (expanded in Appendix~\ref{Appendix - ML model interpretability claim}).

\section{Discussion and Conclusion}
We have presented a data-driven ML approach for determining the joint configuration of concurrent and parallel adapters that maximizes throughput while avoiding request starvation in single-GPU \Gls{LLM}-adapter serving systems. As shown in Table~\ref{table:evaluation_table}, the predicted configurations closely match the real benchmarked optimum, even in heterogeneous scenarios. Notably, simplified and highly interpretable versions of the ML models also achieve competitive performance. Supporting this approach, we developed the first Digital Twin capable of accurately replicating an \Gls{LLM}-adapter serving system, with potential applications beyond generating training data for this work.

\textbf{Limitations}: The key limitation of this study is that the proposed approach, as well as the conducted experiments, targeted only single-GPU configurations. Future work will explore multi-device and multi-node extensions to evaluate performance in realistic production settings and allow comparison with methods such as dLoRA~\cite{wu2024dlora}. In addition, because the approach predicts only the number of concurrent and parallel adapters rather than the allocation of specific adapters, it may perform suboptimally in scenarios with very few adapters that exhibit highly diverse characteristics.


\section{Acknowledgments}
This work has been partially financed by the EU-HORIZON MSCA programme under grant agreement EU-HORIZON MSCA GA.101086248. Also, it has been partially financed by Generalitat de Catalunya (AGAUR) under grant agreement 2021-SGR-00478, by Severo Ochoa Center of Excellence CEX-2021-001148-S-20-3, and by the Spanish Ministry of Science (MICINN), the Research State Agency (AEI) and European Regional Development Funds (ERDF/FEDER) under grant agreement PID2021-126248OB-I00, MCIN/AEI/10.13039/ 501100011033/ FEDER, UE.

\clearpage

\bibliographystyle{IEEEtran}
\bibliography{references}

\clearpage

\appendix

\section{Main LLM-adapter serving overheads}\label{Appendix - Main LLM-adapter serving overheads}
We describe the three main overheads that we encountered when working with adapters, providing more insights into the variations of the targeted optimal configuration. In order to compare the impact of different input/output lengths, we create three synthetic datasets by repeating a single request sampled from the clean ShareGPT dataset at the 25th percentile, mean, and 75th percentile of input/output length: \textit{SmallRequest} (23/27 tokens), \textit{MediumRequest} (250/231), and \textit{LargeRequest} (423/358). We use Llama-2-7B and Llama-2-13B models with LoRA adapters also based on a publicly available adapter~\cite{llama-2-7b-sql-lora-test}.

\subsection{Increased memory usage}
\begin{figure}[!htb]
    \centering
    \includegraphics[width=\linewidth]{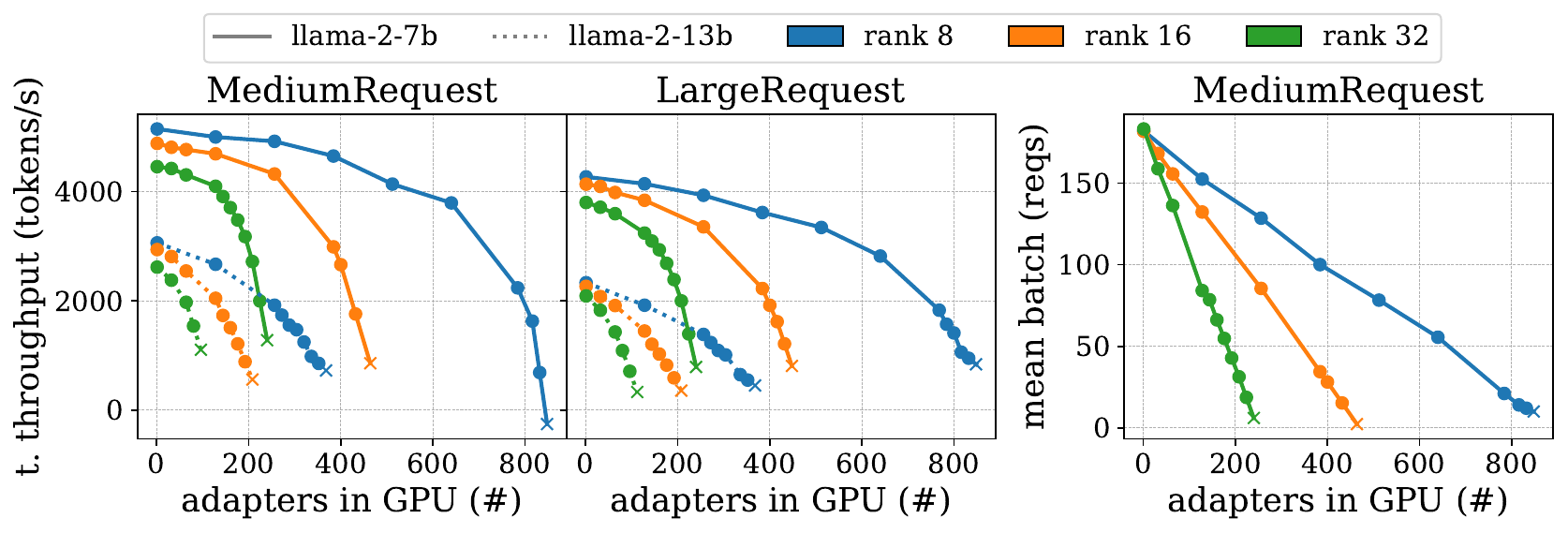}
    \caption{Maximum throughput (left) and batch size (right) evolution as the number of loaded adapters increases, shown for both models, varying adapter sizes/ranks and two datasets. Crosses indicate when no space is left for loading more adapters.} 
    \label{fig:performance_analysis-memory_overhead_full}
\end{figure}

The increased usage of GPU memory for saving adapters' weights reduces the available space for requests' KV cache values, limiting the batch size, and consequently the throughput. In this way, as depicted in left part of Figure~\ref{fig:performance_analysis-memory_overhead_full}, as the number of loaded adapters increases, the maximum achievable throughput also decreases in a somehow exponential pattern. This decrease occurs earlier and happens to be more pronounced with larger models and adapter sizes in accordance with their higher memory usage. Figure~\ref{fig:performance_analysis-memory_overhead_full} also shows the corresponding impact in batch size, which decreases linearly with the number of loaded adapters. This linear trend contrasts with the exponential decline observed in throughput, a difference that is linked to a broader characteristic of \gls{LLM} serving not just adapter serving. As reported in several works, increasing the batch size beyond a certain point leads to diminishing returns in throughput\textemdash a phenomenon known as the throughput plateau~\cite{recasens2024towards, recasens2025mind}\textemdash which corresponds to the leftmost points of each line in the figure.

\textbf{Insight.} Each loaded adapter affects maximum throughput, depending on model, request length, and adapter size\textemdash but this impact fades at large batch sizes due to the throughput plateau.

\subsection{Increased computational workload}
\begin{figure}
    \centering
    \includegraphics[width=\linewidth]{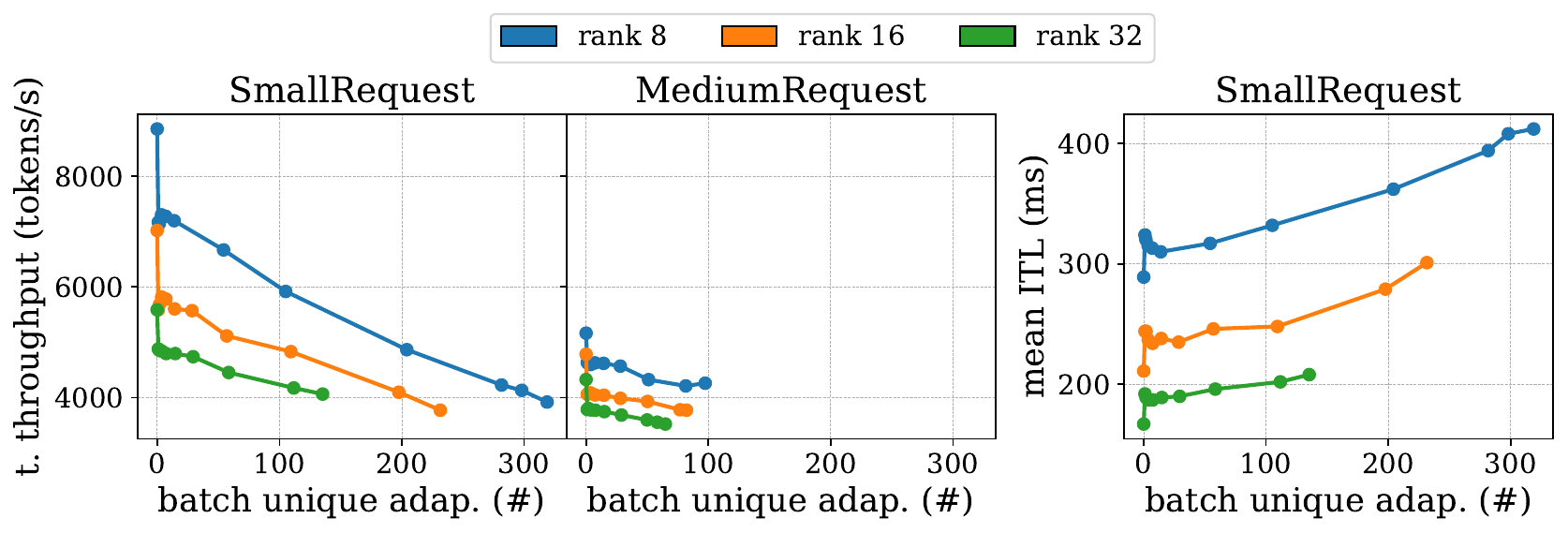}
    \caption{Maximum throughput (left) and \Gls{ITL} (right) as the number of unique adapters in the batch increases, shown for Llama-2-7B, varying adapter sizes/ranks, and two datasets. Lines terminate at the point where the batch size can no longer be increased.}
    \label{fig:performance_analysis-compute_overhead_full}
\end{figure}

Building upon the analysis by Li et al.~\cite{li2024caraserve}, Figure~\ref{fig:performance_analysis-compute_overhead_full} illustrates the impact of increasing the number of unique adapters in the batch on both throughput and \gls{ITL}. Adapter weights introduce overhead to activation computation and GPU cache transfers. The most significant slowdown occurs when moving from zero to one adapter, as this introduces a sequential computation step that cannot be parallelized across requests. Beyond this point, the overhead does not scale in proportion to the number of adapters. While throughput degradation is relatively consistent across datasets, smaller requests can be more affected due to their higher batch sizes, which allow more adapters per batch.

\textbf{Insight.} Serving more unique adapters significantly reduces throughput, bounded by the maximum batch size.

\subsection{Loading time}
\begin{figure}[!htb]
    \centering
    \includegraphics[width=\linewidth]{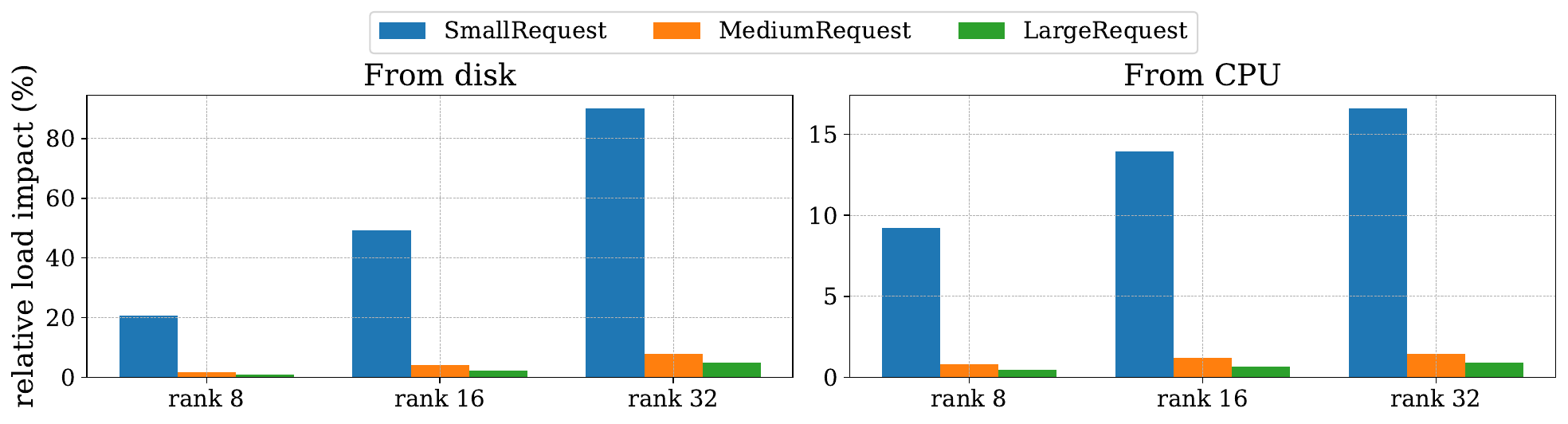}
    \caption{Loading times for varying adapter sizes, shown relative to request latency across the three datasets for Llama-2-7B, and storage type. Request latency is computed as \(TPOT * (output\_tokens - 1)\) where TPOT is the time per output token.}
    \label{fig:performance_analysis-loading_overhead_full}
\end{figure}

Building upon the analysis by Iliakopoulou et al.~\cite{iliakopoulou2024chameleon}, Figure~\ref{fig:performance_analysis-loading_overhead_full} presents the loading time relative to request latency, distinguishing between loading from disk or CPU memory. Larger adapters incur greater overhead, and loading from disk is, on average, 70\% slower than loading from CPU memory. Furthermore, the request length significantly influences the relative impact: for small requests, loading from CPU introduces a latency overhead of 7–16\%, depending on adapter size, whereas for longer requests, this overhead drops below 2\%. Since longer requests require more computation time, the fixed cost of loading becomes comparatively smaller.

\textbf{Insight.} Loading overhead is significant only for short requests and can be largely mitigated by preloading adapters into CPU memory.

\section{S-LoRA case}\label{Appendix - S-LoRA case}
\begin{figure}[!htb]
    \centering
    \includegraphics[width=0.5\linewidth]{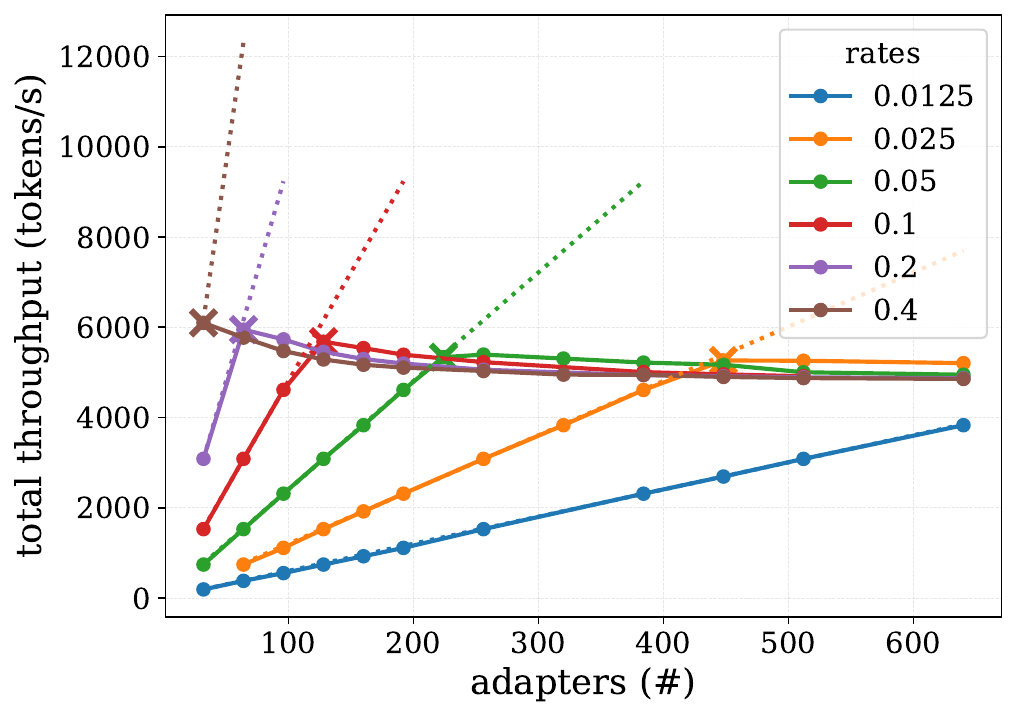}
    \caption{Optimal concurrent adapters when working with S-LoRA (mark with crosses) with varying adapter rates in Llama-2-7B, with rank 32, 250 input tokens and 231 output tokens.}
    \label{fig:performance_analysis-slora}
\end{figure}

We include a brief analysis using the S-LoRA framework~\cite{sheng2024slora} to demonstrate that our problem is not specific to vLLM and can be applied to other frameworks. Figure~\ref{fig:performance_analysis-slora} presents the optimal configuration of concurrent adapters in S-LoRA across different arrival rates. Notably, the decline in throughput as the rate decreases is remarkably modest compared to vLLM, highlighting the relevance of the S-LoRA design. Nevertheless, we still can perceive a 15-20\% decrease in the maximum throughput. Our approach could be handy in identifying these throughput variations and determining the number of adapters at which this maximum is achieved.

\section{Digital Twin extended description}\label{Appendix - Digital Twin extended description}
\begin{figure}
    \centering
    \includegraphics[width=\linewidth, trim=0 5 0 5, clip]{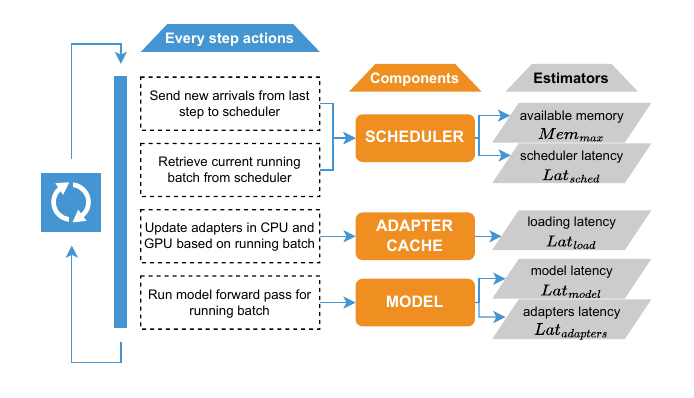}
    \caption{Digital Twin behavior and architecture.}
    \label{fig:digital_twin-architecture_diagram}
\end{figure}

As introduced, the proposed Digital Twin (DT) is an offline emulator that replicates an online \Gls{LLM}-adapter serving system that serves requests from multiple adapters of the same backbone model. Specifically, we try to replicate the widely used vLLM~\cite{kwon2023efficient} framework. As LLMServingSim~\cite{cho2024llmservingsim}, we mirror the "infinite" loop over the running batch of modern LLM serving systems (online batching)~\cite{yu2022orca}, enabling the accurate estimation of key metrics. Figure~\ref{fig:digital_twin-architecture_diagram} illustrates the overall behavior of this loop, the different components involved, and their interactions. The DT is build on modular components, each responsible for simulating a specific aspect of the system, that rely on predictive performance models\textemdash described in Subsection~\ref{subsection: DT - Estimators}\textemdash to predict the time required on the different tasks. Each iteration of the loop follows the same sequence of actions. First, new request arrivals are collected based on the current simulation time and the workload characteristics, and forwarded to the scheduler. The scheduler manages the running batch, removing finished requests and adding new ones. As vLLM, we replicate a \Gls{FCFS} policy with greedy allocation of KV cache and chunked prefill. Once the running batch is updated by the scheduler, it is sent to the adapter cache and model components, which replicate the loading and unloading of adapters with a \Gls{LRU} policy and the model forward pass, respectively.

To run the DT, the expected workload and adapter characteristics\textemdash adapter rates and sizes\textemdash and key server configuration parameters\textemdash mainly the number of adapter slots\textemdash are given. In addition, the DT only needs the expected average and standard deviation of the input and output lengths.

\subsection{Predictive performance models}\label{subsection: DT - Estimators}
Among the five performance models shown in Figure~\ref{fig:digital_twin-architecture_diagram}, \(Mem_{max}\) works an auxiliary estimator that was not included in the main text. It estimates the available GPU memory for storing KV values based on the number of adapter slots and adapter sizes, covering the memory overhead discussed in Appendix~\ref{Appendix - Main LLM-adapter serving overheads}. This estimator is built directly from retrieving the values obtained in that section. The remaining four predictive models predict the latency of each component of the loop, as previously seen in Equation~\ref{eq:digital_twin-latency}. While \(Lat_{load}\) is also derived from direct benchmarking of the loading process across various sizes\textemdash coming from Appendix~\ref{Appendix - Main LLM-adapter serving overheads} results\textemdash, the other three use simple linear models fitted to benchmarking data. \(Lat_{sched}\), representing the scheduler time, is estimated based on the number of running requests (\(R_{running}\)), waiting requests (\(R_{waiting}\)), and an interaction term involving \(R_{waiting}\) and the ratio of adapter slots (\(G\)) to concurrent adapters (\(N\)), reflecting the behavior of the original scheduling algorithm. \(Lat_{model}\), the latency of the base model, is estimated solely from the number of running requests, following prior analyses~\cite{zhang2023shepherd, 10.1145/3341301.3359658}. Finally, \(Lat_{adapters}\), corresponding to the computational overhead discussed in Appendix~\ref{Appendix - Main LLM-adapter serving overheads}, is modelled as an overhead and estimated based on the number of parallel adapters (\(A_{running}\)), in accordance with our results.

\section{Novelty in Digital Twin predictive performance models}\label{Appendix - Novelty in Digital Twin predictive performance}
As outlined in the main text, the predictive performance models in Equation~\ref{eq:digital_twin-latency} can be viewed as either simplifications or modifications of prior work. Since this characterization is somewhat broad, we provide a more detailed explanation here. For the latency of the backbone LLM, our approach does not introduce much novelty: prior studies have employed linear models with respect to the number of concurrent requests (batch size) for generic AI models~\cite{zhang2023shepherd, 10.1145/3341301.3359658}, which is consistent with our benchmark results. For the latency overhead introduced by adapters, Li et al.~\cite{li2024caraserve} estimate it based on the sum or maximum of the adapter ranks present in the batch. However, their formulation did not align with our results, leading us instead to model the latency overhead as a function of the number of parallel adapters, which has a way higher impact than the rank in our results. Finally, regarding scheduler time, to the best of our knowledge we are the first to propose such a model for \Gls{LLM} serving. That said, this formulation is currently limited to vLLM, and further evaluation across other frameworks is required to assess its generality.

\section{Digital Twin extra results}\label{Appendix - Digital Twin extra results}
\begin{figure}[!htb]
    \centering
    \includegraphics[width=\linewidth]{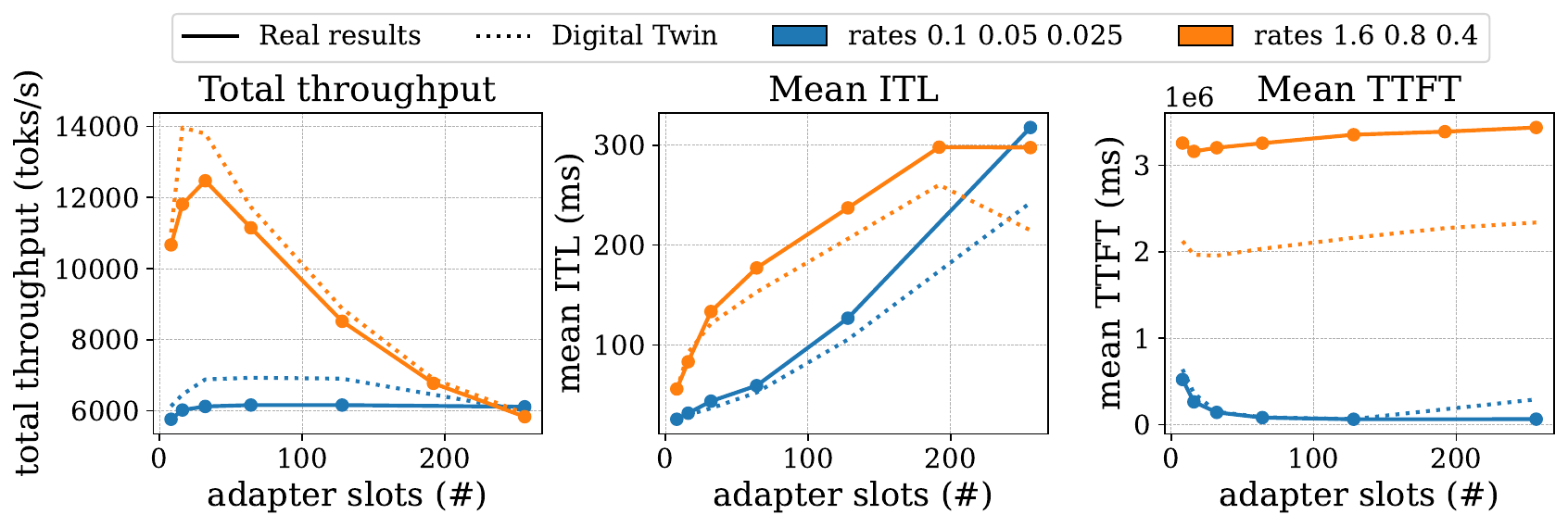}
    \caption{Comparison between DT and real results for throughput, \Gls{ITL} and \Gls{TTFT} for varying adapter slots and rates when serving 256 concurrent adapters of ranks 8 and 16 on Qwen-2.5-7B model.}
    \label{fig:evaluation-dt}
\end{figure}

Figure~\ref{fig:evaluation-dt} visually illustrates the differences between the Digital Twin and real system results for one of the evaluated combinations. Consistent with the table results, the DT accurately estimates throughput and \Gls{ITL}, while \Gls{TTFT} shows larger deviations. Additionally, the figure highlights that \Gls{ITL} estimation worses with a higher number of adapter slots, and \Gls{TTFT} becomes less accurate under higher arrival rates.

\section{ML model extra results}\label{Appendix - ML model extra results}
Table~\ref{table:evaluation_table_interpretable_model} expands the results of the right portion of Table~\ref{table:evaluation_table} for all tested model types.

\begin{table}[!htb]
    \renewcommand{\arraystretch}{1.2}
    \centering
    \begin{tabularx}{\columnwidth}{c l *{3}{>{\centering\arraybackslash}X}}
        \toprule
        Model & Estimator & Concurrent adapters (\%) & Adapter slots (\%) & Time (ms) \\
        \midrule
        \multirow{7}{*}{\rotatebox{90}{Llama-3.1-8B}} 
            & LinearRegression   & 40.61 & 19.53 & 0.04 \\
            & BayesianRidge      & 40.59 & 17.54 & 0.25 \\
            & PLSRegression      & 40.61 & 17.38 & 0.04 \\
            & RFRegressor        & \textbf{0.05} & \textbf{6.73} & 0.13 \\
            & RuleFitRegressor   & 6.36 & 14.89 & 1.80 \\
            & FIGSRegressor      & 0.14 & 14.06 & \textbf{0.03} \\
            & \textit{Tree**} & 3.70 & 10.86 & \textbf{0.10} \\
        \midrule
        \multirow{7}{*}{\rotatebox{90}{Qwen-2.5-7B}} 
            & LinearRegression   & 38.59 & 23.70 & \textbf{0.03} \\
            & BayesianRidge      & 38.57 & 24.57 & \textbf{0.03} \\
            & PLSRegression      & 38.56 & 23.62 & \textbf{0.03} \\
            & RFRegressor        & \textbf{1.01} & \textbf{7.15} & 0.15 \\
            & RuleFitRegressor   & 1.88 & 11.76 & 2.03 \\
            & FIGSRegressor      & 1.11 & 14.97 & \textbf{0.03} \\
            & \textit{Tree**} & \textbf{1.01} & 12.41 & 0.09 \\
        \bottomrule
    \end{tabularx}
    \caption{Expanded results of the ML model for predicting the optimal joint configuration (concurrent adapters and adapter slots). All values represent the SMAPE difference from the benchmark maximum values, except time (reported in milliseconds), which corresponds to the summed inference time across the two output predictions.}
    \label{table:evaluation_table_interpretable_model}
\end{table}

\section{ML model interpretability claim}\label{Appendix - ML model interpretability claim}
\begin{figure}[ht]
    \centering
    \begin{subfigure}[t]{0.95\textwidth}
        \centering
        \includegraphics[width=\linewidth]{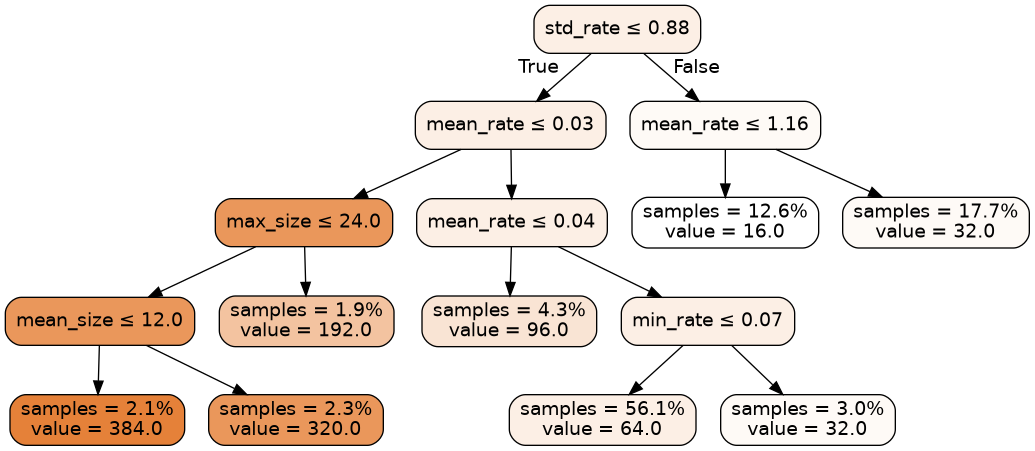}
        \caption{Simplified RFRegressor}
    \end{subfigure}%
    \hfill
    \begin{subfigure}[t]{0.95\textwidth}
        \centering
        \begin{minipage}{\linewidth}
        {\scriptsize
        \begin{tabular}{l}
            \\
            1) \(\widetilde{rate} \le 0.88 \;\&\; \overline{rate} \le 0.03\;\&\; max(size) \le 24.\;\&\; \overline{size} \le 12. \rightarrow 384\) \\
            2) \(\widetilde{rate} \le 0.88 \;\&\; \overline{rate} \le 0.03 \;\&\; max(size) \le 24. \;\&\; \overline{size} > 12. \rightarrow 320\) \\
            3) \(\widetilde{rate} \le 0.88 \;\&\; \overline{rate} \le 0.03 \;\&\; max(size) > 24. \rightarrow 192\) \\
            4) \(\widetilde{rate} \le 0.88 \;\&\; \overline{rate} > 0.03 \;\&\; \overline{rate} \le 0.04 \rightarrow 96\) \\
            5) \(\widetilde{rate} \le 0.88 \;\&\; \overline{rate} > 0.03 \;\&\; \overline{rate} > 0.04 \;\&\; min(rate) \le 0.08 \rightarrow 64\) \\
            6) \(\widetilde{rate} \le 0.88 \;\&\; \overline{rate} > 0.03 \;\&\; \overline{rate} > 0.04 \;\&\; min(rate) > 0.08 \rightarrow 32\) \\
            7) \(\widetilde{rate} > 0.88 \;\&\; \overline{rate} \le 1.16 \rightarrow 16\) \\
            8) \(\widetilde{rate} > 0.88 \;\&\; \overline{rate} > 1.16 \rightarrow 32\) \\
        \end{tabular}}
        \end{minipage}
        \caption{Extracted rules}
    \end{subfigure}
    
    \caption{(Top) Simplified RFRegressor used to predict the number of adapter slots for the Qwen-2.5-7B model. (Bottom) The same model expressed as a set of eight rules,, where \(\overline{x}\) and \(\widetilde{x}\) denote the mean and standard deviation of \(x\), respectively.}
    \label{fig:interpretability_claim}
\end{figure}

We simplify the best-performing model, the RFRegressor, by reducing it to a single decision tree (parameter \textit{n\_estimators}) with a maximum depth of six (parameter \textit{max\_depth}) and enforcing a minimum of ten samples per leaf (parameter \textit{min\_samples\_leaf}). This reduction yields a highly interpretable model that retains competitive performance, as shown in Table~\ref{table:evaluation_table_interpretable_model} under the \textit{Tree**} row. An example of such a tree, trained to predict the number of adapter slots, is presented in Figure~\ref{fig:interpretability_claim}, which also illustrates the set of simple rules in the form “condition 1 AND condition 2 … → RESULT” that can be extracted from the estimator. Equivalent models were generated for the concurrent adapters output, requiring no more than 30 rules. These rule-based representations provide an interpretable decision source for production systems while offering a lightweight, efficient, and practical solution for real-world deployments.

\end{document}